\documentclass[manuscript]{aastex}

\usepackage{graphicx, amsmath}

\shorttitle{Shaping the Brown Dwarf Desert}
\shortauthors{Jumper, P.H. and Fisher, R. T.}

\begin{document}


\title{Shaping the Brown Dwarf Desert: Predicting the Primordial Brown Dwarf Binary Distributions from Turbulent Fragmentation}  

\author{Peter H. Jumper and Robert T. Fisher}
\email{robert.fisher@umassd.edu}
\affil {University of Massachusetts Dartmouth, 285 Old Westport Road, N. Dartmouth, MA, 02747-2300}


\begin{abstract}

The formation of brown dwarfs (BDs) poses a key challenge to star formation theory. The observed dearth of nearby ($\leq 5$ AU)  brown dwarf companions to solar-mass stars, known as the brown dwarf desert, as well as the tendency for low-mass binary systems  to be more tightly-bound than stellar binaries,  have been cited as evidence for distinct formation mechanisms for brown dwarfs and stars. In this paper, we explore the implications of the minimal hypothesis that brown dwarfs in binary systems originate via the same fundamental fragmentation mechanism as stars, within isolated, turbulent giant molecular cloud cores.  We demonstrate analytically that the scaling of specific angular momentum with turbulent core mass naturally gives rise to the brown dwarf desert, as well as wide brown-dwarf binary systems. Further, we show that the turbulent core fragmentation model also naturally predicts that very low-mass (VLM) binary and BD/BD systems are more tightly-bound than stellar systems.  In addition, in order to capture the stochastic variation intrinsic to turbulence, we generate $10^4$ model turbulent cores with synthetic turbulent velocity fields to show that the turbulent fragmentation model accommodates a small fraction of binary brown dwarfs with wide separations, similar to observations.   Indeed, the picture which emerges from the turbulent fragmentation model is that a single fragmentation mechanism may largely shape both stellar and brown dwarf binary distributions during formation.




\end{abstract}



\section {Introduction}

A key unresolved topic in star formation is the nature of the brown dwarf desert. The desert refers to the observed relative absence of brown dwarf companions to solar mass stars within a distance of 5 AU \citep {gretherlineweaver06}.  Alternatively, Marcy and Butler 2000 described the desert by noting that 0.5\% of primary stars with masses $> 0.5 M_{\odot}$ had brown dwarf companions within orbits of 3 AU.  Several  theories have been proposed to explain the desert, including  ejection from low-order multiple stellar systems \citep{reipurthclarke01}, disk fragmentation \citep {riceetal03} followed by inward disk migration and destruction \citep {armitagebonnell02}, subsequent ejection \citep {bateetal03, goodwinwhitworth07}, and the separation of stellar and BD forming objects from distances of ~70 AU from the protostar, with the former migrating inward and accreting additional mass \citep{sw09}. 

Furthermore, it has been argued that the observed differences in  the orbital properties of very low mass (VLM) and brown dwarf systems from those of stellar systems supports different formation mechanisms for stars and brown dwarfs. In addition to the brown dwarf desert,  observations of  VLM and brown dwarf binary systems  reveal that these systems typically have  narrower separations of $a < 20$ AU \citep{burgasseretal07}. Moreover,  binary brown dwarfs typically have a significantly higher cutoff on their minimum binding energies than stellar systems \citep{kroupaetal03, kroupa2011, parkergoodwin11}.  In this viewpoint, a different mechanism for the formation of brown dwarfs and stars (e.g., disk fragmentation versus molecular cloud core fragmentation) may give rise to a break in the IMF across the substellar boundary \citep {thieskroupa07} as well as distinct set of binary properties for BDs and stars  \citep{kroupa2011}.  An N-body study following the evolution of very low mass binary systems, with total system mass $\leq$ $\ 0.2 M_{\odot}$ has, however, demonstrated that such tight, hard systems cannot be disrupted even in the densest stellar clusters, which suggests that these systems must be formed at birth \citep {parkergoodwin11}.

Moreover,  more recent theoretical work has emphasized limitations in the idealized thermodynamic treatment of protostellar disks in  earlier star formation models \citep {rafikov05}.  Additional theoretical models of the collapse of turbulent cores into protostellar disks further showed that the protostellar disks would likely remain stable to brown dwarf formation if the effects of stellar irradiation, which were neglected in early computational models, were properly accounted for \citep{matznerlevin05, krumholz06}.   However, Matzner and Levin 2005 qualified that warm, highly turbulent cores, such as those forming massive stars, could still result in disk fragmentation forming brown dwarfs \citep{matznerlevin05}, while Whitworth \& Stamatellos 2006 argued that brown dwarf formation could occur within 100 AU of protostars. Subsequent radiation hydrodynamic simulations, taking into account stellar irradiation feedback, have shed light on brown dwarf formation process and have found that irradiated protostellar disks are more gravitationally stable and produce fewer brown dwarfs than models which neglected stellar irradiation \citep {krumholzetal07, bate09, offneretal09, bate12}, consistent with theoretical predictions. As a counterpoint, Stamatellos, Whitworth, and Hubber 2011 argue that episodic accretion and/or radiative feedback could mitigate the effects of irradiation, thereby promoting disk fragmentation.  State-of-the-art numerical simulations including the effects of radiative feedback have begun to yield  samples of up to 200 stars and brown dwarfs \citep {bate11},  but cannot yet provide enough systems to make statistically-significant statements about the nature of the brown dwarf desert. Consequently, the turbulent core formation model, in which brown dwarfs form from low-mass, turbulent-pressure confined cores, remains a viable though largely unexplored model \citep {padoannordlund04}.


On the observational side, numerous observations are consistent with the view that brown dwarfs and stars may share a common formation mechanism \citep {luhmanetal07}. In particular, like classical T-Tauri systems, young brown dwarfs exhibit accretion infall signatures in H-alpha and near-infrared excesses indicative of accretion disks \citep {muzerolleetal03}. The accretion rate within these  substellar systems is significantly lower than observed in classical T-Tauri systems, with the mass accretion rate scaling as the square of the system mass \citep {mohantyetal05}.  However, as disk fragmentation \citep{sw09} and ejection \citep{bate09, bate11} may also produce brown dwarfs with accretion disks, this evidence is not conclusive.

In addition, observations have highlighted the existence of a small number of widely separated, soft VLM and BD-BD binaries in star-forming regions. These systems include the wide young binary BD system 2MASS J11011926-7732383AB at a separation of 242 AU \citep {luhman04},  wide young BD system Oph 162225-240515 at a separation of  243 AU \citep {rayjayivanov06, closeetal07},   wide young BD binary UScoCTIO 108 at a projected separation of 670 AU \citep {bejaretal08}, and wide young BD binary FU Tau A and B \citep {luhmanetal09} at a separation of 800 AU.  In the field, observations have located the VLM binary DENIS-J055146.0-443412.2 at a separation of over 200 AU \citep {billeresetal05} field VLM binary LEHPM 494/DENIS-P J0021.0-4244  at a separation of 1800 AU \citep {caballero07}, and   VLM binary 2M0126AB at a projected separation of 5100 AU \citep {artigauetal07, artigauetal09}.

Such widely-separated VLM and BD-BD systems are very rare and account for no more than 1\% - 2\% of all systems, but nonetheless play a crucial role in constraining models of star formation. In particular, wide separations of young BD/BD and BD/stellar binaries significantly wider than a typical protostellar disk size of $100$ AU pose a significant challenge for disk fragmentation models of brown dwarf binary formation if formed through interactions in the disk \citep{whitworthstamatellos06}; however, the ejections and pairings of brown dwarfs from disks may allow some BD/BD systems to form \citep{sw09}. The current VLM record-holder is 2MASS J12583501+4013083 and 2MASS J12583798+4014017, a very wide binary with both components near the BD mass limit, with a projected separation of 6700 AU \citep {radiganetal09}. Such very soft systems will be softened and ultimately disrupted by dynamical interactions with field stars and giant molecular clouds, but as \citet {radiganetal09} note,  their estimated lifetimes, based upon standard binary evolutionary models \citep{weinbergetal87} can still be $\sim$ several Gyr. 

In light of the current state-of-the-art of computational simulations, a theoretical framework of a simplified model will help to illuminate the key physics of the brown dwarf desert and the scaling behavior of other low mass stellar systems.  Specifically, in this paper, we build upon and extend a turbulent fragmentation model of binary formation within isolated turbulent cores  \citep{fisher04}(hereafter F04), and explore the implications of this model for the formation of low-mass and brown dwarf binary systems. In the turbulent core model, the mass and angular momentum of binaries are initially established by the mass and angular momentum of an isolated, parent turbulent core. Consequently, the mass and angular momentum of the turbulent core gives rise to the initial, or ``primordial''  distributions of both stellar/BD and BD/BD binaries immediately subsequent to fragmentation, and prior to any subsequent dynamical evolution, including  possible effects of competitive accretion and gravitational torques from the surrounding cluster protostellar members.  Significantly, as we will re-derive, the turbulent core model predicts that the specific angular momentum of turbulent cores scales with the mass of the core to the 3/4 power \citep {burkertbodenheimer00}. {\it We will demonstrate that the scaling of specific angular momentum with turbulent core mass in turn naturally produces a brown dwarf desert, as well as wide brown-dwarf binary systems.} Finally, we will also demonstrate that the turbulent fragmentation model also naturally predicts that low-mass binary and BD/stellar systems are more tightly-bound than stellar systems.

We should note that the isolated core model is an idealization of the complex dynamics of giant molecular clouds. Historically, the concept that GMCs are highly inhomogeneous, clumpy structures with smaller, denser cores within these clumps was motivated by noting that if the densities of CO clumps reflected the mean density of the GMC, the GMC extinction would exceed that observed by over an order of magnitude \citep {blitzshu80}. However, clumps within GMCs are far more  complex than the ballistically-moving isolated objects originally envisioned by Blitz and Shu; current models suggest that they are the densest gaseous regions within highly-complex flows governed by supersonic turbulent dynamics \citep {padoanetal97, padoannordlund02}. Moreover, recent three-dimensional  hydrodynamical numerical simulations of supersonic turbulent giant molecular clouds demonstrate that GMC cores are dynamical entities, and continue to accrete from a network of parent filaments, even as their embedded stars are in the process of formation \citep {ballesterosparedes03, offneretal10, hansenetal11}. 

The isolated core idealization, however, remains a fruitful one because it allows us to explore the physics of turbulent fragmentation for statistically-significant numbers of binary systems. Moreover, the isolated turbulent core fragmentation model allows us to  capture the role which turbulence plays in giving rise to the angular momentum of the parent gas distribution \citep {burkertbodenheimer00}.  F04 demonstrated that this angular momentum distribution of parent GMC cores, in turn, plays an important role in  establishing  binary periods and separations. In particular,  the isolated turbulent core model robustly agrees with a number of observed properties, both of prestellar cores \citep {burkertbodenheimer00, matznerlevin05}, and of pre-main sequence and field binaries (F04). Specifically, it agrees with observational data on the width of the binary period distribution. Moreover, both the model and observations exhibit an anticorrelation of binary period and mass ratio, and a positive correlation of binary period and eccentricity. The turbulent core model also predicts that low-mass model binaries originate within lower-mass turbulent cores with less angular momentum, and also naturally yield narrower binaries than stellar-mass systems, in accord with observation \citep{closeetal03}.  This broad agreement with the predictions of the isolated turbulent core model for stellar systems motivates an extension of the model into the substellar regime. 


This paper is organized as follows. Section 2 discusses estimated scaling relationships that provide simplified predictions of our model's results.  Section 3 focuses upon the methodology of our semi-analytic model for the formation of both stellar and substellar binaries from turbulent isothermal GMC cores, with a particular emphasis on advancements made to our F04 model to accommodate substellar systems. Section 4 presents the results of our model. Lastly, section 5 presents our discussion of the results and conclusions.

\section{Scaling Relations for Turbulent Fragmentation}

\subsection{Derivation}

Turbulence in the core provides it with a net angular momentum \citep {burkertbodenheimer00}. Intuitively, as Landau originally suggested long ago \citep {landaulifshitz59, davidson09}, we can understand the net angular momentum generated by a turbulent power spectrum consistent with Larson's linewidth-size relation \citep {larson81} by realizing that although the net angular momentum induced by  the numerous small-scale turbulent modes will tend to cancel out, the few large-scale turbulent modes which fit within the core will tend to result in a small net angular momentum. Turbulence is inherently a stochastic physical process, and different realizations of a turbulent velocity field  will endow a core with differing levels of angular momentum. However, the resulting net mean angular momenta of models of turbulent cores are nonetheless consistent with the low mean rotation rates implied by the linewidth gradients of cores mapped in NH$_3$ \citep {goodmanetal93, jijinamyersadams99}.  

\citet {burkertbodenheimer00} showed that because most of the angular momentum endowed by turbulence is generated at the scale of the core, the mean specific angular moment of a population of turbulent cores could be reasonably estimated even under the simplifying assumption of uniform core rotation.  We therefore assume uniform rotation to derive estimates for the specific angular momenta of model cores.  Expanding upon this, we derive estimates for binary properties, including the scaling of semimajor axes, periods, and binding energies with system mass from the isolated turbulent core fragmentation model. These estimates capture the essential physical description of turbulent fragmentation, which we will subsequently elaborate upon in \S \ref {methodology} in more detailed calculations, taking into account a fuller description of the stochastic variation inherent in different realizations of  turbulence, as well as the inhomogeneous, turbulent GMC background within which individual cores are embedded. 

In our derivation of the scaling relations, we assume that the core is a critical Bonnor-Ebert sphere \citep{ebert55, bonnor57} with a density distribution modeled  using an analytic lowered power law approximation \citep{natarajanlyndenbell97}.  We calculate the moment of inertia $I$ and the gravitational potential energy $\Omega$ of this core as  $ I = cM_{\rm core}R_{\rm core}^2$ and  $\Omega = -dGM_{\rm core}^2/R_{\rm core}$, where $c\approx 0.34$ and $d\approx 0.55$ are constants  numerically determined for a critical Bonnor-Ebert sphere, in terms of the core mass $M_{\rm core}$ and radius $R_{\rm core}$.  We take the parameter $\beta =cR_{\rm core}^3\omega^2/(2dGM_{\rm core}) $ to describe the ratio of rotational energy to gravitational binding energy of the core, and find the specific angular momentum of the critical Bonnor-Ebert core  $J_{\rm core}/M_{\rm core}$ scales as the square root of the mass and radius of the core; $J_{\rm core}/M_{\rm core} =  \sqrt{2cd \beta GM_{\rm core}R_{\rm core}}$.  We combine the Larson turbulent velocity dispersion relation with an exponent of $1/2$, $\sigma = 1.10 $ km s$^{-1} L$(pc$)^{0.5}$, with the condition that the core is in virial balance; in terms of the virial parameter, $\alpha = 5 \sigma^2 R/(GM)$, $\alpha \sim 1$.  Consequently, we find that $J_{\rm core}/M_{\rm core} \propto M_{\rm core}^{3/4} \propto  R_{\rm core}^{3/2} \ $\citep{larson81, leungetal82, myers83}. The latter scaling reflects the increase of line width with increasing size, and is the same scaling reported by \citet {burkertbodenheimer00}.  These lead to a scaling estimate of the specific angular momentum with the mass of the core:

%
\begin{equation}
{J_{\rm core}\over M_{\rm core}}  = 2.6 \times 10^{20}\left({\alpha \over 1.3}\right)^{1/4} \left({\beta \over 0.02}\right)^{1/2}\left({M_{\rm core}\over M_{\odot}}\right)^{3/4} {\rm cm}^2\ {\rm s}^{-1}
\label {jovermscaling}
\end{equation}

Crucially, $lower$-$mass$ turbulent cores in virial balance naturally have a lower specific angular momentum than more massive cores. This scaling of the specific angular momentum with mass has profound consequences for binary properties.   
While most of the mass and angular momentum in a core is carried away during the star formation process and does not end up in the final binary, we can describe the fractions of mass and angular momentum transferred from the core to the binary system in terms of a star formation efficiency, $\epsilon_* = M/M_{\rm core}$, and an angular momentum efficiency, $\epsilon_J = J/J_{\rm core}$, where $M = M_1 + M_2$ and $J$ are the total mass and angular momentum of the binary system, with a primary mass $M_1$ and a companion mass $M_2$.  Studies have suggested that the star formation efficiency is fairly constant over a wide range of formation conditions, with a typical value of $0.3$ \citep{alvesetal07}.  Less information is known about the angular momentum efficiency, but F04 demonstrated that the stellar period distribution could be reproduced by a constant value of $\epsilon_J$ for a given model star formation efficiency $\epsilon_*$.  Thus, we use the two efficiencies to derive the scaling of the system specific angular momentum with total mass from the core scaling:

\begin{equation}
{ J  \over  M}= 3.37 \times 10^{19} \left({\epsilon_J  \over 0.016}\right)\left({0.30\over \epsilon_*}\right)^{7/4}\left({\alpha \over 1.3}\right)^{1/4}\left({\beta \over 0.02}\right)^{1/2}\left({M \over M_{\odot}}\right)^{3/4} {\rm cm}^2\ {\rm s}^{-1}
\end{equation}

We may then use this scaling to estimate the typical periods $P$ and semimajor axes $a$ of binary systems : 

\begin{equation}
P = 159  \left({\epsilon_J  \over 0.016}\right)^3\left({0.30\over \epsilon_*}\right)^{21/4}\left({\alpha \over 1.3}\right)^{3/4}\left({\beta \over 0.02}\right)^{3/2}\left({M \over M_{\odot}}\right)^{1/4}{1 \over (1-e^2)^{{3 \over 2}}  } { (1+q)^6 \over q^3  } {\rm  days}
\end{equation}



\begin{equation}
a = 0.57 \left({\epsilon_J  \over 0.016}\right)^2\left({0.30\over \epsilon_*}\right)^{7/2}\left({\alpha \over 1.3}\right)^{1/2}\left({\beta \over 0.02}\right)\left({M \over M_{\odot}}\right)^{1/2} {1 \over (1-e^2) }  {(1+q)^4\over q^2  } {\rm AU},
\end{equation} 
where $e$ is the eccentricity of the system and $q = M_2 / M_1$ is the mass ratio.

As seen above, the  period and semimajor axis scale weakly with mass, to the $1/4$ and $1/2$ powers respectively.  Moreover, the mass ratio of the system, $q$, is crucial for shaping the primordial distributions, as most of the angular momentum that is transferred from the core to the binary system will be associated with the orbit of the companion.        

To illustrate the significance of the mass ratio concretely, consider two $1 \ M_{\odot}$ systems: one a stellar binary with $q = 1$ and the second a BD/stellar binary with $q = 0.04$. For simplicity, we assume  both have the mean eccentricity of a thermal distribution, $e = 2/3$.  With our fiducial scalings, the semimajor axes of the stellar and BD/stellar binary will be 16 AU and 750 AU, respectively.  The wider separation of the BD/stellar binary systems is typical of the brown dwarf desert.  Thus, the desert naturally arises in the turbulent fragmentation model primarily as a mass ratio effect.

Furthermore, consider a $0.16 \  M_{\odot}$  binary brown dwarf system with $q = 1$, which will have a semimajor axis of 6.6 AU with our fiducial scalings.  Consequently, the turbulent fragmentation model also naturally predicts that VLM and binary brown dwarf systems will be narrower than stellar binary systems, as they are formed within low-mass turbulent cores with lower specific angular momenta than stellar-mass turbulent cores.  Moreover, these lower-mass systems will have nearer-equal mass ratios than stellar systems; this trend toward equal mass ratios further favors narrow-binary brown dwarf systems.

We may expand upon our semimajor axis estimates to construct the turbulent fragmentation model prediction for the scaling of the minimum binding energy with the system mass.  A system of mass $M$ has binding energy of approximately $E_{\rm bind} = GM_1M_2/a \propto  M^{3/2}(q^3/(1+q)^6)$.  Such a binary is most weakly bound when it has a brown dwarf companion; these systems establish the minimum binding energies of a binary with total mass $M$.  For a companion dwarf with a much lower mass than the primary, $M_2 << M_1$, this scaling will approximately follow $E_{\rm bind, min}  \propto  M^{-3/2}$.  As an example, let us consider a $0.10 \ M_{\odot}$ and a $1.00 \  M_{\odot}$ system, each with a $0.01 M_{\odot}$ companion.  We find that the former system is bound $\approx 22$ times more tightly than the latter.  Our result is similar to the conclusion derived by Close et al. 2003 that VLM and binary brown dwarf systems tend to be 10 - 20 times more tightly bound that solar mass systems.           


\section{Methodology}
\label {methodology}

The turbulent fragmentation model describes the formation of binary systems from turbulent giant molecular cloud cores.  The mass and angular momentum due to turbulence in these parent cores set the mass and angular momentum in the resulting system, determining its orbital properties.  This paper's key addition to the methodology developed in F04 is the relaxation of the assumption of a fixed core edge pressure. Instead, we consider cores embedded within an isothermal, supersonically turbulent giant molecular cloud, with a wide range of edge pressures that naturally result in a wide range of critical Bonnor-Ebert masses.  These include dense, low mass cores that produce substellar binary systems alongside stellar systems formed from more massive cores. 


Conceptually, our model of turbulent cores is similar to one recently proposed by \citet {padoannordlund11}, which is based upon an earlier model of the IMF \citep {padoannordlund02}. Both models assume that background supersonic isothermal turbulence establishes core edge pressures and that the core masses are uncorrelated with these edge pressures. However, \citet {padoannordlund11} consider the core mass function (CMF), as observed in a star-forming region, and assume that the formation of cores is distributed uniformly in time. The ongoing core accretion within their model leads to a {\it local} star-formation efficiency, defined in terms of the instantaneous core mass and the final stellar mass, greater than unity. In contrast, our model considers only the final core masses when determining binary properties, which are fixed by the {\it global} star-formation efficiency $\epsilon_*$. Furthermore, both models predict a prevalence of pressure-confined cores at substellar core masses less than the local Bonnor-Ebert mass. These cores are gravitationally-stable and will not form stars or brown dwarfs; thus we reject any pressure-confined cores generated in our model.   

Some of the core mass and angular momentum will be lost,  through either stellar winds and outflows \citep {matznermckee00}, or magnetic braking \citep {basumouschovias94}, respectively, during the star formation process.  Thus, only a fraction of the original core mass and angular momentum remains in the binary star system at the time of formation.  These fractions are described by two parameters,  the star formation and angular momentum efficiencies,  $\epsilon_* = M_{\rm tot} / M_{\rm core}$ and  $\epsilon_{\rm J} = J_{\rm tot} /  J_{\rm core}$ respectively.  Observational studies suggest that the star formation efficiency is $\epsilon_* \approx 0.3$,  independent of mass \citep{alvesetal07, andreetal10}.  This is also in agreement with Machida and Matsumoto's  MHD simulation results, which suggest that cores have $0.3 \leq \epsilon_* \leq 0.5$ \citep{machidamatsumoto11}.   

To determine an angular momentum in the parent core, and the binary system through the angular momentum efficiency, we impose a Gaussian random turbulent field upon the core.  As in F04, we set this efficiency so that the distribution of stellar binary systems reflects observations (see Table 3.1).  For $\epsilon_* = 0.3$, we find that $\epsilon_J = 0.016$ in our model.





In F04, masses were drawn according to the Kroupa initial mass function (IMF)  \citep{kgt91}.  However, this IMF had a lower cutoff at 0.08 $M_{\odot}$ and thus excluded the possibility of modeling brown dwarfs in the substellar range.  In this paper, we adopt the Chabrier 2005 IMF \citep{chabrier05}, which extends the lower cutoff of the MF down to the deuterium-burning mass limit of 0.01 $M_{\odot}$. The Chabrier 2005 IMF differs from the Chabrier 2003 IMF \citep{chabrier03}, which overpredicts the relative abundance of brown dwarfs to stars.   By shifting the peak of its lognormal segment to better account for the relative numbers of brown dwarfs and stars, the Chabrier 2005 IMF predicts the formation of one brown dwarf between the masses of $0.03 \ M_{\odot}$ and $0.08 \  M_{\odot}$ for every four stars with masses $\leq 1\ M_{\odot}$.  When discussing the Chabrier IMF throughout this paper, we will be referring to the Chabrier 2005 IMF.        


The Chabrier IMF is a piecewise-defined function, obeying a lognormal distribution for masses $ \leq 1\ M_{\odot}$ and a power law for masses $> 1\ M_{\odot}$\citep{chabrier05}.  The power law regime of the Chabrier IMF has an index of $-2.35 \pm 0.3$ \citep{chabrier05}, in accordance with the famous Salpeter IMF index of $-2.35$ \citep{salpeter55}.

To draw masses from the Chabrier IMF, we must first express it as a probability distribution function of mass (in $M_{\odot} $),  $m$.  Chabrier provides his IMF, normalized in terms of volumetric quantities, as a function of the common logarithm of mass, $\log{m}$ :

\begin {equation}
{\xi}\ (\log{m}) =
\begin{cases}
A  \exp{\left[-{(\log{m}- \log{m_c})}^2 / {\left(2\times \sigma^2\right) }\right]} & \text{if } m \leq  m_o, \\
Bm^{-x} & \text{if } m > m_o.
\end{cases}
\end {equation}

 $A$ and $B$ are prefactors, $m_c$ gives the location of the peak of the lognormal distribution, $\sigma$ describes the width of the lognormal distribution, $-x-1$ is the power law index, and $m_o$ is the location of the break in the IMF.  Here, these parameters have values $A = 0.093$, $B = 0.041$, $m_c = 0.20$, $\sigma = 0.55$, $x =-1.35 \pm 0.3$, and $m_o = 1$ \citep{chabrier05}.



Conversion between ${\xi}\ (\log{m})$ and ${\xi}\ ({m})$ is conducted by multiplying the former by $1/(m\ln{10})$ \citep{scalo86, chabrier03}.   Therefore, the probability distribution function (PDF) of the IMF may be expressed in terms of two additional prefactors $A_*$ and $B_*$ as follows:

\begin {equation}
{\rm PDF}\ (m) =
\begin{cases}
\left(A_*/\ln{10}\right)  m^{-1} \exp{\left[-{(\log{m} - \log{m_c}})^2 / {\left(2\times \sigma^2\right) }\right]}  & \text{if } m \leq  m_o, \\
\left(B_*/\ln{10}\right)  m^{-x-1} & \text{if } m > m_0.
\end{cases}
\end {equation}

We must normalize the PDF  to unity under the constraint that the IMF is continuous at $m_o = 1$. We then calculate the values of $A_*$ and $B_*$ necessary to satisfy both the continuity and normalization of the PDF :

\begin{equation}
A_* = \ln{10} \left[\sqrt{\pi \over 2}\ \sigma \ln{(10)} \ {\rm erfc} \left[  \log{m_c} - \log{m_o}  \over  \sigma \sqrt{2}  \right] +{1 \over x}\exp{\left[-\left(\log{m_o} -\log{m_c} \right) ^2 \over 2\sigma^2 \right]} \right]^{-1}
\end{equation}
\begin{equation}
B_* = A_*m_o^x  \exp{\left[-\left(\log{m_o} -\log{m_c} \right) ^2 \over 2\sigma^2 \right]}
\end{equation}
Substituting, we find that $A_* \approx 0.724$ and $B_* \approx 0.323$.   

%
%

We calculate the cumulative distribution function (CDF) for the Chabrier  IMF and utilize the inverse transform sampling method to draw masses from it for our model :
\begin {equation}
{\rm CDF}\ (m) =
\begin{cases}
A_* \sigma \sqrt{\pi / 2}\ {\rm erfc} \left[(\log{m_c} - \log{m} ) / {(\sigma\sqrt{2})} \right] & \text{if } m \leq  m_o, \\
C  + (B_* / \ln{10})( m_o^{-x}/x) ( 1 - \left(m/m_o\right)^{-x} ) & \text{if } m > m_o.
\end{cases}
\end {equation}
Here $ C \approx 0.896$, the value of the CDF at $m = 1 \ M_{\odot}$, ensures the continuity of the function across the break in masses.  

%
%


We assume that binary masses are uncorrelated and thus draw both masses in each system independently from the IMF.  This assumption works well for low stellar masses, but breaks down at high masses;  O and B type stars tend to be paired with near equal mass stars \citep{sana09}.  Therefore, we account for the limits of random pairing by restricting all stars  in our model to $\leq 2 \ M_{\odot}$.  Furthermore, by drawing the masses as an input in this manner, our model assumes that each of its stars and brown dwarfs primordially form as part of a binary.  Additionally, as our model is primordial, we will not directly capture any dynamical evolution of the binary population as cluster evolution causes soft binaries to become softer and hard binaries become harder \citep{weinbergetal87}.

Once we have drawn the primary and companion masses, we set the parent core mass by $M_{\rm core} = M_{\rm tot}/\epsilon_*$, where $M_{\rm tot}$ is the total binary system mass.   We assume that the core is described by a turbulently-supported Bonnor-Ebert sphere \citep{ebert55, bonnor57} at or exceeding the critical edge density at which the core becomes Jeans unstable.  Once we determine a core edge density, we utilize a Bonnor-Ebert sphere density profile to provide the core's overall density structure.  Combining this information with a Gaussian random turbulent velocity field, we may calculate a realization of the core angular momentum.

We assume a GMC model with mass $ M = \ 10^5 \ M_{\odot}$ in virial equilibrium.  To do so, we take the $2500$ $M_{\odot}$ GMC model assumed by \citep {krumholzmckee05} and scale it to $10^5$ $M_{\odot}$ under an isothermal temperature of 10 K.  Using this model GMC, we calculate a dispersion velocity of $\sigma_{cl} = 4.0$ $\rm km \ s$$^{-1}$ and then use the mean density relationship derived by Krumholz and McKee 2005 to find an mean mass density, $\rho_{\rm cl}$:  
\begin{equation}
\rho_{\rm cl} ={375 \over 4\pi}{\sigma_{\rm cl}^6 \over \alpha_{\rm vir}^3 G^3 M_{\rm cl}^2},
\end{equation}    
From this, we derive a mean model GMC mass density of $4.8\times 10^{-21}$ g cm$^{-3}$.  Using standard mass fractions of molecular hydrogen and helium, this corresponds to a number density of  $\approx 1300$ cm$^{-3}$, which provides the expectation value of our lognormal distribution.  This allows us to draw the core edge densities from the distribution.

Supersonic isothermal turbulence produces a lognormal distribution of pressures (and thus densities) within GMCs \citep{padoanetal97}.  Thus, we draw the edge densities of our model cores from such a distribution.  However, as we have drawn our stars and brown dwarfs from the IMF, we have effectively assumed that the cores in our model must collapse.   We accept edge densities that make the cores Jeans-unstable; any lower densities resulting in Jeans-stable cores are rejected and redrawn. 

Padoan and Nordlund give a log-normal probability distribution function for number density $n$:
\begin{equation}
p(n)dn = {1  \over (2\pi \sigma^2)^{1/2} n}\exp{\left[-{1\over 2}\left({\ln{(n)} - {\overline {\ln{(n)}}  }\over \sigma } \right)^2\right]}
\end{equation}
 based on an mean number density of $1$  $ {\rm cm^{-3}}$, which results in $\overline{\ln{n}} = -\sigma^2/2$ \citep{padoannordlund04}.  For different mean densities, the relationship between the mean of the distribution $\mu$ and $\overline{\ln{n}}$ becomes  $\overline{\ln{n}} =\ln{(\mu)} -\sigma^2/2$. We utilize the mean number density determined previously ($\approx 1300$ cm$^{-3}$), substitute this into the number density PDF and utilize the inverse transform sampling method as before to draw edge densities from the CDF of the Padoan and Nordlund log-normal distribution.

 Padoan and Nordlund describe $\sigma^2$ in terms of two parameters, $b$, which reflects the degree to which the model's turbulent forcing is solenoidal or compressive, and $M_S$, the turbulent Mach number, such that $\sigma^2 = \ln{(1 + b^2 M_S^2)}$ \citep{padoannordlund04, federrath10}.  Models assuming solenoidal forcing have included proposals of $b = 0.26$ \citep{kritsuk07} and $b = 1/3$ \citep{price2011}.  The actual nature of the forcing is unknown, but it is likely an intermediate between solenoidal and compressive,  similar to the original proposal of $b = 0.5$ \citep{padoanetal97}, which we will adopt for our model.



In addition to both of the binary system's masses, the parent core's edge density, and Gaussian random turbulence, we must also draw the eccentricity of the final binary system to complete our orbital calculations.  This value is drawn from the thermal distribution, $f(e) = 2e$, which has a mean value of $2/3$.  Together, these values yield the binary's mass, angular momentum, eccentricity, and mass ratio, which are then used to calculate the binary's period and semimajor axes:    


%
\begin{equation}
P = \left( {2\pi \over G^2 }  \right) \left( {J^3 \over M^5 }\right) {1 \over (1-e^2)^{{3 \over 2}}  } { (1+q)^6 \over q^3  }
\end{equation}

\begin{equation}
a = {1 \over G} \left( {J \over M }\right)^2 {1 \over (1-e^2) } {1\over M  } {(1+q)^4\over q^2  }
\end{equation} 

We repeat these procedures for each binary, generating $10^4$ model systems.  



While we assume that the majority of our systems will produce core fragmentation, we acknowledge that disk fragmentation may be a possibility in some systems.  However, we estimate that the contribution of disk fragmentation  to the overall fragmentation in this  low-mass range is small.  To assess the contribution of disk fragmentation, we consider \citet{offneretal10},  who derived a criteria to predict binary or multiple formation through disk fragmentation. \citet {offneretal10} describe their models through two parameters, $\xi$ and $\Gamma$, which represent the dimensionless accretion rate onto the disk/stellar system (compared to the isothermal rate), and the rotational rate of the  core, as determined by the ratio of the orbital period of the infalling gas to the accretion time scale.  \citet{offneretal10} demonstrated that disk fragmentation occurs for typical values of $\Gamma$ (0.001 - 0.01) at values of $\xi \approx$ 2-3.   We estimate the fraction of systems with $\xi > 2$ as a function of mass; this occurs for  approximately $ 14\%$ of systems at 0.16 $M_{\odot}$ and approximately $32\%$ of systems at 1.00 $M_{\odot}$.   This analysis, which approximates the accretion rate onto a single stellar/disk system, places an upper-bound to the expected disk fragmentation in our model in this low-mass range.

\section {Results}
\label{results}

\subsection{The Primordial Brown Dwarf Desert}

The key feature of the brown dwarf desert is the relative absence of brown dwarf companions within 3-5 AU of primary stars of around a solar mass compared to stellar companions at similar distances to such primaries \citep{marcybutler00, gretherlineweaver06}.  Studies have proposed upper limits between 0.5\% and 1.0\% on stars exceeding 0.5 $M_{\odot}$  having brown dwarf companions within these distances \citep{marcybutler00, gizisetal2001, gretherlineweaver06}.  Likewise, approximately $11\pm 3$\% of solar-type stars have stellar companions within this regime \citep{gretherlineweaver06}.  In our model, we drew 3709 primary stars with $M > 0.5 M_{\odot}$ out of a sample of $10^4$ systems.  The prevalence of brown dwarf companions within narrow systems was within the limits implied by observations, with 7, or 0.19\%, such companions within 3 AU and 12, or 0.32\% within 5 AU of their solar-type primaries.  Far more stellar companions were found within this regime, with 232, or 6.26\% within 3 AU and 307, or 8.28\%, within 5 AU.  Thus, our model tends to somewhat underpredict the frequency of narrow stellar companions but nonetheless accommodates a brown dwarf desert.   

\begin{figure}
	\plotone{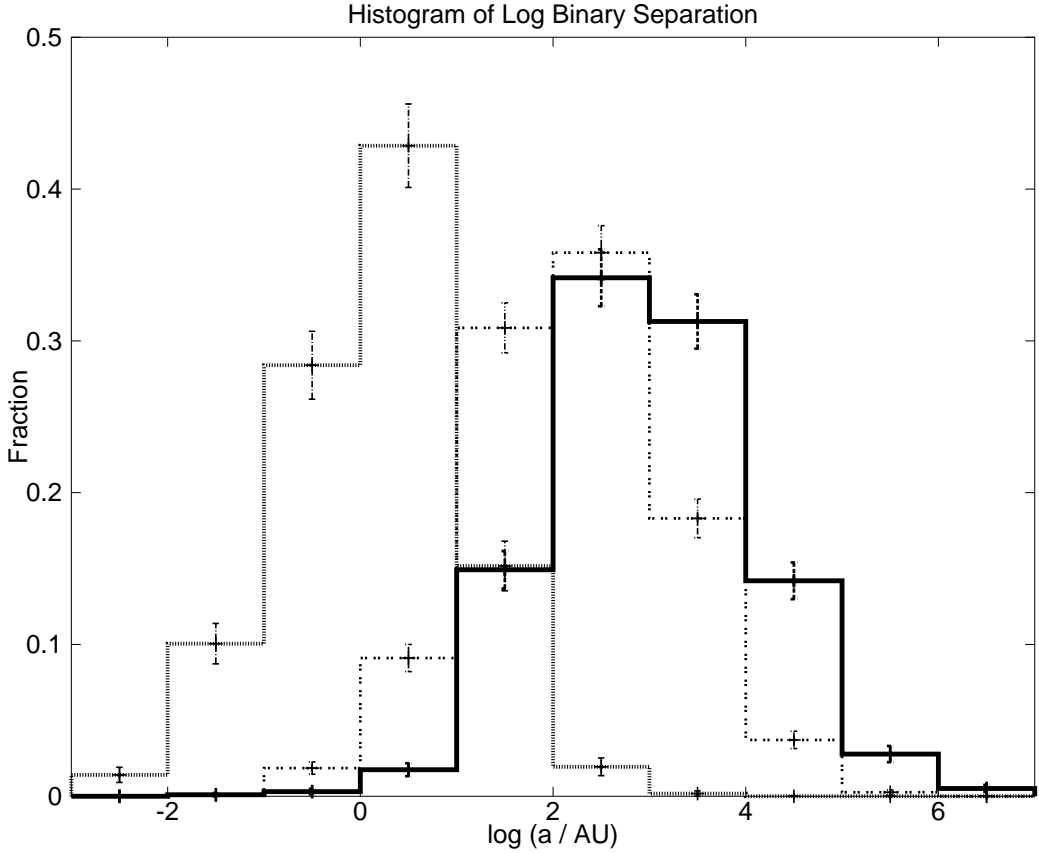}
	\caption{Probability distribution of log semimajor axes for 972 model stellar/BD binary systems with primary masses $\geq 0.5 M_{\odot}$ (solid line), 567 model BD/BD binary systems (dashed line), and 1131 model systems with primary masses $\geq 0.5 M_{\odot}$ and companion masses between $0.08 M_{\odot}$ and $0.20 M_{\odot}$ (dotted line).  For all systems, $\epsilon_* = 0.3$ and $\epsilon_J = 0.016$.  Error bars have been calculated with Poisson statistics.}
   \label{fig:BDBinary3}
\end{figure}

The distribution of semimajor axes for solar/BD systems is shown in Figure \ref{fig:BDBinary3}, compared against both the distributions of solar/VLM systems and BD/BD systems.  The solar/BD distribution details  all model brown dwarf companions to  primary stars between 0.5 $M_{\odot}$ and 2 $M_{\odot}$ under the efficiencies of $\epsilon_* = 0.30$ and $\epsilon_J = 0.016$, a sample of 972 systems.  Among this population, narrow systems remain scarce.  Only 21 of 972 such systems, or 2.2\%, have separations $\leq \ 10 \ {\rm AU}$, while 166, or 17\%, have separations $\leq \ 100 \ {\rm AU}$.  

Moreover, the peak of the semimajor axis distribution is located on the order of 1000 AU, as seen in Figure \ref{fig:BDBinary3}, demonstrating our model brown dwarfs' preference for wide separations.  This is further supported by additional measures of central tendency, including a median of the semimajor axes of $\approx 911 \ {\rm AU}$.  Thus, our results are consistent with the conclusions of Gizis et al. \citep{gizisetal2001}, who suggested that brown dwarf companions may be common at distances greater than 1000 AU.  Interestingly, this result is also similar to the disk fragmentation model's prediction of a peak at approximately 800 AU \citep{sw09}, even though different mechanisms were used to produce this result.  


Additionally, Figure \ref{fig:BDBinary3} shows that the model solar/BD systems characteristic of the desert are far less likely to be found at narrow separations than model solar/VLM systems with companions masses between $0.08 M_{\odot}$ and $0.20 M_{\odot}$, further demonstrating the dearth of narrow brown dwarf companions compared to stellar companions at these separations.    

\begin{figure}
	\plotone{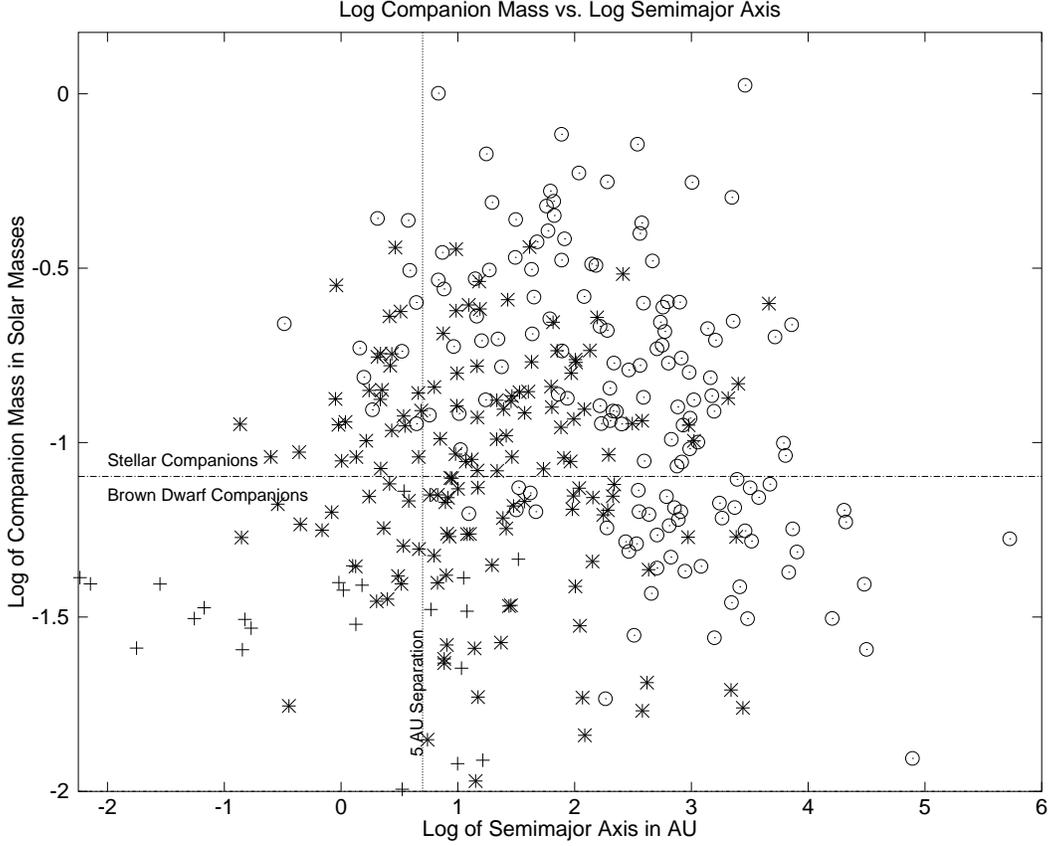}
	\caption{Companion masses in solar masses, on a log scale, versus system semimajor axes in AU for 400 model systems. The systems are randomly selected out of each of three populations such that their relative numbers reflect the model's distribution.  Circles depict systems with primary stars between $0.5 M_{\odot}$ and 2.0 $M_{\odot}$, asterisks indicate systems with primary stars between $0.08 M_{\odot}$ and 0.5 $M_{\odot}$, and pluses denote systems with brown dwarf primaries with masses between $0.01 M_{\odot}$ and $0.08 M_{\odot}$.  A horizontal line demarcates systems with brown dwarf companions from those with stellar companions, while a vertical line marks semimajor axes of 5 AU.}
	\label{fig:CompanionSeparation}
\end{figure}

The brown dwarf desert is also demonstrated in Figure \ref{fig:CompanionSeparation}, which plots the log of the companion mass versus the log of the semimajor axis in AU for companions to solar primaries ($0.5 M_{\odot} \leq M \leq 2.00 M_{\odot}$, circles), low mass primaries ($0.08 M_{\odot} < M < 0.5 M_{\odot}$, asterisks) and BD primaries (pluses) for 400 randomly chosen model systems.  Dashed lines are drawn for a companion mass of $0.08 M_{\odot}$ and a semimajor axis of 5 AU, indicating the regime of the brown dwarf desert in the lower-left corner, in which there is dearth of brown dwarf companions to solar stars.   Instead, these companions are preferentially found at separations of 100s or 1000s of AU.  Additionally, the figure shows that the model has simultaneously produced BD/BD systems with predominantly narrow separations.   

Crucially, the current model only considers the primordial orbital parameters that result from the initial formation of its binary systems.  Therefore, the production of a brown dwarf desert suggests that turbulent fragmentation may explain the desert primordially, prior to the influence of further evolutionary effects.  This contrasts with the previous evolutionary models explaining the emergence of the brown dwarf desert from later dynamical effects, such as migration or ejection, as noted in the introduction.  Such evolutionary mechanisms continue to shape the binary distribution well past  formation, but the binary distributions themselves are largely established at birth \citep {parkergoodwin11}.

\subsection{Brown Dwarf - Brown Dwarf Binaries}

Binary brown dwarf systems preferentially form at narrow separations.  In a review paper, Burgasser et al. 2007 reported that 93\% of VLM systems have separations less than 20 AU.  Additionally, they indicated that the peak of the binary brown dwarf semimajor axis distribution is located between 3 AU and 10 AU, with a mean of 4.6 AU.  Similarly, Close et al. 2003 reported that the peak of this distribution was located at $\approx 4 \ {\rm AU}$.  Our model produces similar results, as shown in Figure \ref{fig:BDBinary3}'s depiction of BD/BD binaries, which provides data for 567 binary brown dwarfs systems at $\epsilon_* = 0.3$ and $\epsilon_J = 0.016$.  We find that 512, or 90.3\% of our model systems have separations less than or equal to 20 AU, with 469, or 82.7\% of systems within the narrower limit of 10 AU.  Our distribution appears to peak at a somewhat lower value than those reported by Burgasser et al. and Close et al, peaking in the vicinity of its median of $\approx 1.63 \ {\rm AU}$.  


Furthermore, the mean of our distribution, $\approx 15.4 \ {\rm AU}$, initially appears to contrast with the observational results of Burgasser, Close, and coauthors.  However, due to limited statistics, most observational samples do not capture many of the widest binary brown dwarf systems, which may have semimajor axes in the hundreds to thousands of AUs, with a current record 6700 AU \citep {radiganetal09}.  In contrast, our model has captured a long tail in the  primordial semimajor axis distribution, including 12 systems with separations in excess of 100 AU, extending to a maximum of $a \approx 3481 \ {\rm AU}$.  This tail results from turbulence in cloud cores; different realizations of the turbulence in the the same core may result in greatly different angular momenta, and in turn orbital separations.  This inherent stochastic nature of turbulence gives rise to unusually wide binary brown dwarfs, which increase the mean separation of our sample.  Moreover, the primordial nature of the tail is significant, as soft, or weakly bound, binaries tend to become softer and more widely separated over time.  Therefore, while would not expect to find a primordial separation rivaling the current record holder in the field, we could expect that a time evolution of the model systems would produce more wide systems and some systems with separations of a few thousand AU.  Thus, the turbulent fragmentation model can accommodate both the existence of a thin tail of wide systems as well as the more typical, narrower systems.  


\subsection{Binding Energies}

Observational studies have found that very low mass and binary brown dwarf systems tend to be bound more tightly than stellar systems, demonstrating that the minimum binding energy for VLM and BD binaries is roughly 10-20 times higher than that for solar mass systems \citep{closeetal03, burgasseretal07}.  In Figure \ref{fig:BindingEnergy}, we compare our model results against a compilation of previous observational results presented in Burgasser et al. 2007's Figure 6 (191 systems) and the vlmbinaries database (94 systems).  Circles represent our model systems in Figure \ref{fig:BindingEnergy} (285 systems), while stars represent the observational results.   Additionally, we draw a dashed line representing the approximate scaling of the minimum binding energy derived in \S2, $E_{\rm bind, min} \propto M^{-3/2}$.  We base this line from the binding energy predicted for a $1.01$ $M_{\odot}$ system with a $1.00$ $M_{\odot}$ primary and a  $0.01$ $M_{\odot}$ secondary ($q = 0.01$) by our scaling estimates, $1.6 \times 10^{40}$ ergs.  

As seen in  Figure \ref{fig:BindingEnergy}, our model results produce an trend of increasing minimum system binding energy with decreasing total system mass, consistent with the expectation presented by observation. Both the observations and our model include a few far more weakly bound outliers.  Moreover, the minimum binding energies presented by both our data points and our scaling estimate line are similar to those found observationally and to each other.    More generally, there is a wide overlap of the model and observational results throughout the parameter space.  While there appear to be concentrations of observed binaries not produced in the model, it is important to remember that the model sample was drawn uniformly from our results, without preference for a particular regime.  In constrast, the observational results are composed from several studies, many of which were investigating specific stellar populations.  This produces  a bias with concentrations of observed data at these regions.   Ultimately, the similarities between our model and the observational results suggest that the physics of turbulent fragmentation during primordial formation may play a key role in establishing the observed trend in minimum binding energies.       
     

\section {Conclusions}

The turbulent fragmentation model  provides an alternative to the hypothesis that brown dwarfs form a distinct population with a separate formation mechanism from stars and that the IMF is discontinuous across the substellar mass boundary \citep {thieskroupa07}.  In particular, the brown dwarf desert, the greater binding energies of VLM and BD/BD systems compared to solar mass systems, and the result that ``one BD is produced for every 4-6 formed stars" have all been used as evidence to argue in favor of such distinct populations \citep{thieskroupa07}. In this paper, we have demonstrated that, when properly scaled to account for both the turbulent linewidth as well as turbulent core edge pressure, the predictions of turbulent fragmentation can accommodate observed systematic trends in binary properties. Therefore, given that a turbulent fragmentation model may also explain these binary brown dwarf properties in terms of a single core fragmentation mechanism, we suggest that these binary properties are not strong evidence for separate formation mechanisms for brown dwarfs. Indeed, the picture which emerges from the turbulent fragmentation model is that a single fragmentation mechanism largely shapes both stellar and brown dwarf binary distributions during formation. Well-understood, subsequent dynamical interactions, both within the nascent stellar cluster, and with field stars and GMCs will continue to evolve the binary distributions \citep {weinbergetal87}, but these interactions alone cannot account for the observed binary distributions of VLM and BD binaries, as recent N-body models have demonstrated \citep {parkergoodwin11}.

\begin{figure}
    \plotone{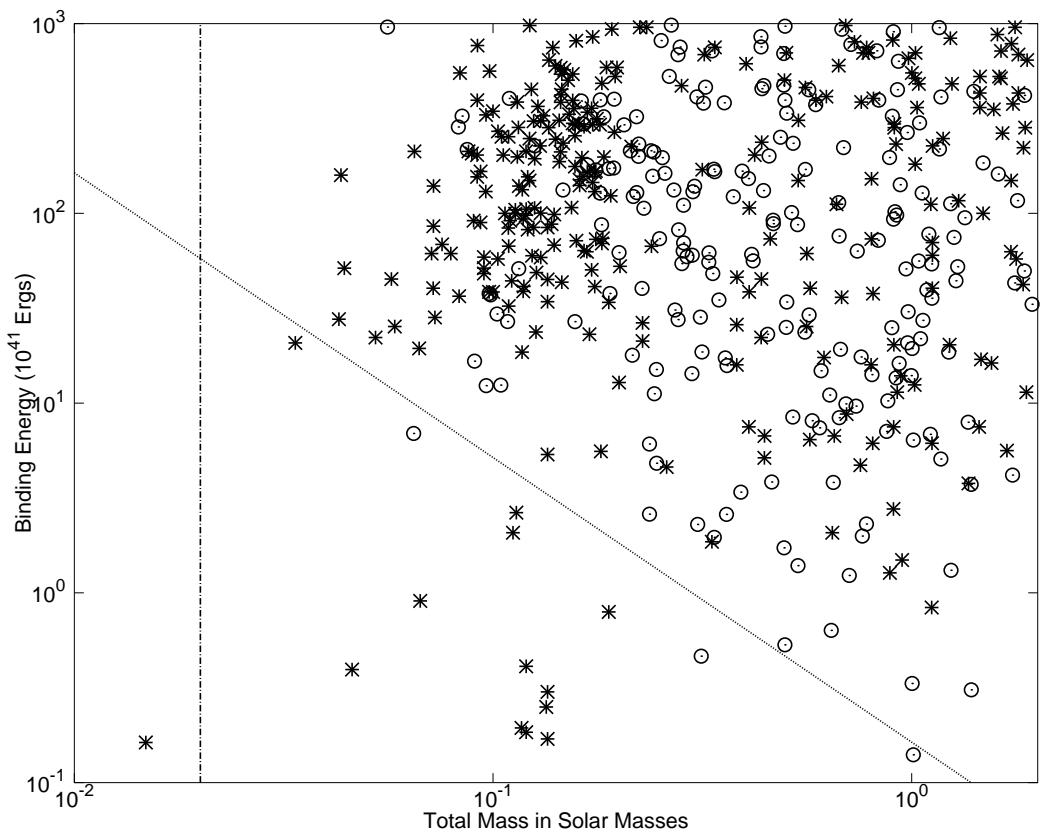}
    \caption{Binding energy versus total system mass for 285 observed systems (191 from Burgasser et al. 2007, 94 from vlmbinaries.org, represented by asterisks), and an equal number of model systems (circles).  The dashed line represents the approximate scaling of $E_{\rm bind, min} \propto M^{-3/2}$.  The solid line represents the minimum mass model binary system.}  
    \label{fig:BindingEnergy}
\end{figure}

One deficiency of the turbulent fragmentation model presented here is that it assumes equal binarities for stars and brown dwarfs, whereas observations have demonstrated  that the overall field binarities for VLM stars and brown dwarfs are lower than those of G dwarfs  \citep{dm91,fischerandmarcy92, closeetal02}.  Our current model addresses only the primordial binary distributions and consequently lacks sufficient depth to address this issue in its entirety. However, a general outcome of turbulent fragmentation is that the most weakly-bound systems preferentially include those with brown dwarf or VLM companions.  Consequently,  we conjecture that the softening and eventual disruption of these most weakly-bound systems will result in a decreased binarity fraction from stellar to brown dwarf masses.  Other authors, evolving VLM systems, have come to a similar conclusion that the brown dwarf binary fraction must have been higher at birth than is now observed \citep {parkergoodwin11}. There is some evidence for this trend in Figure \ref{fig:BindingEnergy}, which demonstrates that several brown dwarf binaries are significantly softer than turbulent fragmentation predicts for the primordial distribution. These ultrasoft BD binaries in the field may be the result of subsequent softening of initially soft primordial BD binaries. However, more work on the evolution of these binaries in larger clusters is necessary in order to quantify the evolution of the binary fractions themselves. 

One key future test of the turbulent fragmentation model predictions for brown dwarf binaries will be the development of numerical simulations of giant molecular clouds in virial equilibrium which can produce statistically large ($N > 10^4$) samples of binaries.  In addition to allowing for a closer comparison to observations in the field, such models may allow further insight into the problems of the BD and VLM binarity fractions and the relative frequencies of BD/BD and BD/stellar systems.  Such simulations will be challenging, but may be within the capability of the next-generation of petascale supercomputers. 

 In addition, the turbulent fragmentation model prediction that low-mass stellar and brown dwarf binaries will have a higher binding energy than solar mass binaries may have broader application to our understanding of the core mass function and the origin of the IMF.  In particular, \citep {padoannordlund11} have recently described how supersonic isothermal turbulence shapes the observable pre-stellar phase of turbulent low-mass cores in principle. In practice, however, it will be extremely challenging to distinguish in both observation and simulation the truly pre-stellar cores from the numerous unbound turbulent transient density fluctuations, which dominate the core mass spectrum at very low masses. However, because the lowest-mass binaries also fragment directly from the  lowest-mass turbulent cores, the turbulent core fragmentation model suggests that the turbulent pre-stellar core phase can be traced in well-established properties of the stellar binary field distribution. This connection between turbulent cores and binaries may thereby allow a direct and relatively clean window into the physics which shapes the turbulent core mass spectrum and, in turn, the initial mass function.

\acknowledgements

The authors acknowledge discussions with Stella Offner, Kaitlin Kratter, and David Ribeiro. RF
acknowledges research support from NSF Grant CNS-0959382 and AFOSR
DURIP Grant FA9550-10-1-0354.  PJ acknowledges support from NSF Grant DMS-0802974.

\clearpage


\begin{table}
\begin{tabular} {|l    |l  |l  |l  |l  |l  |l  | l | }
\hline 
$\epsilon_*$ & $\epsilon_J$ & $\log {\bar {P}_d}$ & $\log {\sigma_{P_d}}$ & $\bar {a} /$ AU  & $\bar {j}$ (cm$^2$ s$^{-1}$) &  $\bar {\mathcal {M}}$ \\
 \hline 0.3  & .016 & 4.9  & 1.6 & 33 & $4.8 \cdot 10^{20}$ & .88  \\
\hline
\end{tabular}
\label{resultstable}

\caption{Distribution of parameters for all stellar systems without brown dwarfs at $\epsilon_* = 0.3$.  As in F04, $\epsilon_J$ is set to fit the median log period of stars.}
\end{table}

 \vfill\eject

\vfill\eject

\vfill\eject

\vfill\eject

\vfill\eject
	
\end{document}